\documentclass[12pt,letterpaper]{article}
\usepackage[utf8]{inputenc}
\usepackage[T1]{fontenc}
\usepackage{amsmath}
\usepackage{amssymb}
\usepackage{graphicx}
\usepackage{url}
\usepackage{bbm}
\usepackage{tocloft}
\usepackage{tabularx}
\usepackage{amscd}
\usepackage{amsthm}
\usepackage[normalem]{ulem}
\usepackage{xcolor}

\usepackage{algpseudocode}

\title{Climate aware lending allocation under NGFS scenarios - A Monte Carlo approach}
\date{ }
\author{Marina Palaisti}

\begin{document}
	\maketitle 
	
	\section*{Abstract}
	
	\paragraph{Purpose:}	
	\textcolor{black}{This paper examines climate-sensitive credit risk and lending allocation over a five-year horizon by mapping NGFS short-term transition and physical-risk scenarios into obligor default probabilities.}
	
	\paragraph{Methodology/Approach:}
	\textcolor{black}{NGFS shocks enter mean-reverting log-production dynamics, yielding analytical terminal default probabilities. Monte Carlo simulation with covariance-based PCA estimates joint losses, while constrained nonlinear optimization accounts for exposure-dependent default risk and compares alternative lending policies.}
	
	\paragraph{Findings:}
	\textcolor{black}{Under a synthetic 1,000-obligor portfolio, mean five-year terminal default probabilities range from 13.88\% (HWTP) to 13.95\% (DAPS). Relative to conventional allocation, climate-aware lending reduces expected credit losses by 0.55--0.71\% under HWTP, SWUC, and DiRe, with negligible improvement under DAPS. Tail risk is sensitive to dependence assumptions, and fixed-PD allocation generates markedly greater concentration.}
	
	\paragraph{Research Limitations:}
	
	\textcolor{black}{Results are conditional on NGFS scenario inputs and synthetic borrower calibration. Sectoral shocks cannot fully represent within-sector differences in emissions, geography, technology, or adaptation, motivating future institution-specific validation.}
	
	\paragraph{Practical Implications:}
	The framework helps banks link NGFS scenarios to sector-level credit risk and portfolio loss outcomes. It can support ICAAP and climate stress testing by highlighting exposures most sensitive to climate risk, adding a forward-looking overlay to existing models.
	
	\paragraph{Originality/Value:}
	\textcolor{black}{The study advances climate-aware credit allocation by linking NGFS sectoral projections to analytical obligor-level default probabilities and endogenous lending decisions. It provides a unified framework for examining how climate-related credit risk influences lending allocations, expected losses, and portfolio concentration, supported by quantitative policy comparisons and robustness analysis.}

	\section*{Introduction}

	\textcolor{black}{Physical climate hazards and the transition to a low-carbon economy affect borrowers through operational disruption, policy changes, technological adjustment, and shifting demand. These channels influence firm value and credit risk over bank capital-planning horizons.}

\textcolor{black}{Credit-risk models calibrated on historical defaults may miss shocks from new climate policies, technologies, and physical hazards. NGFS scenarios allow these forward-looking risks to be traced through borrower creditworthiness and portfolio exposures, supporting ICAAP and stress testing.}

	\textcolor{black}{We link NGFS sector shocks to obligor log-production dynamics, analytical terminal default probabilities, joint-loss simulation, and lending optimization with exposure-dependent default risk. The framework compares creditworthiness, lending income, concentration, and losses across climate scenarios.}

	\textcolor{black}{The study makes three principal contributions. First, it incorporates time-dependent NGFS scenario projections into a structural credit-risk model, translating sector-specific transition and physical-risk information into adjustments to firm production dynamics and terminal default probabilities. This allows climate-related risks to be assessed at the obligor level while preserving the heterogeneity of borrower characteristics and exposures. Second, it examines the interaction between credit risk and lending decisions by accounting for the dependence of default probabilities on the amount of capital allocated to each obligor. The resulting nonlinear optimization problem is evaluated alongside a conventional lending benchmark and a linear approximation based on fixed default probabilities, providing a basis for assessing the economic implications of climate-aware capital allocation. Third, the study evaluates the sensitivity of default risk and portfolio outcomes to alternative dependence structures, parameter calibrations, scenario-transmission assumptions, and diversification constraints. Analytical default-probability calculations, Monte Carlo validation, and systematic robustness experiments provide complementary assessments of the numerical behavior and reliability of the framework.}

	\subsection*{\textcolor{black}{Related Literature and Regulatory Context}}
	
	\textcolor{black}{Recent research has advanced climate stress testing by linking climate scenarios to financial exposures, while identifying important limitations in the transmission from economic shocks to credit outcomes. Reinders et al.~\cite{reinders2025} highlight insufficient sectoral granularity, omitted transmission channels, and the limited treatment of financial feedback effects. Bourgey et al.~\cite{bourgey2025} develop a structural approach to climate-related credit portfolio losses using dimensionality reduction. These contributions strengthen scenario-based risk measurement but leave scope for examining how climate-adjusted default risk influences lending allocation decisions.}

	\textcolor{black}{Existing NGFS-based structural approaches provide an important point of comparison. Battiston et al.~\cite{battiston2025} examine scenario-contingent corporate valuation and credit risk through CLIMACRED, while Ge et al.~\cite{ge2024} assess banks' transition-risk exposures using a Merton-based framework. These studies establish connections between climate pathways, firm creditworthiness, and portfolio exposure. Building on these foundations, the present study integrates scenario-adjusted structural default probabilities with endogenous lending optimization, allowing climate-related changes in creditworthiness to influence portfolio allocation decisions.}

	\textcolor{black}{Empirical evidence further illustrates the heterogeneity of climate-risk transmission. De Bandt et al.~\cite{debandt2025} document substantial variation and measurement uncertainty in estimated banking-sector climate effects. Covi et al.~\cite{covi2025} show that banks adjust lending in response to climate policy, with differences related to existing transition-risk exposures and cross-border relationships. These findings motivate an analysis that accounts for heterogeneous obligor characteristics and scenario-dependent credit risk. Our numerical framework provides a controlled setting for examining how these factors influence lending allocation and portfolio performance.}

	\textcolor{black}{
		Recent developments in explainable artificial intelligence also offer complementary approaches to credit-risk assessment. Machado et al.~\cite{machado2025} demonstrate the use of machine learning and explanatory techniques with observed banking data, while the BIS Project Gaia~\cite{bisgaia2024} illustrates how AI can improve the extraction and comparability of climate-related disclosures. Such methods may support hybrid frameworks combining empirical prediction with structural scenario analysis. Their potential flexibility must nevertheless be balanced against model uncertainty, interpretability, and validation requirements. The present framework prioritizes an explicit, economically interpretable climate-to-credit transmission mechanism.}
	
	\textcolor{black}{The regulatory literature emphasizes additional requirements for implementation. IFRS~9~\cite{ifrs9} incorporates forward-looking economic information into expected credit losses, but requires probability-weighted loss measurement and credit-risk assessments that extend beyond a single scenario-conditioned terminal default probability. The Basel Committee~\cite{bcbs2022} emphasizes governance, risk identification, and supervisory oversight, while recent EBA guidelines~\cite{eba2025management,eba2025scenario} strengthen expectations concerning ESG risk management and environmental scenario analysis.}

	\subsection*{Problem Formulation}

	\textcolor{black}{Consider a financial institution allocating a lending budget $K$ across $N$ obligors. Let $x_i$ denote the loan allocated to obligor $i$, and let $\eta^i$ denote its lending rate. Under climate scenario $s$, the terminal probability of default is
		\[\mathrm{PD}^i_T(x_i;s)= \mathbb{P}\!\left(V_T^i(x_i;s)\leq L^i(T)\right),\]
		where $V_T^i$ is the firm's value at maturity and $L^i(T)$ is its default threshold. Since the allocated loan amount affects firm production through a leverage-related drift adjustment, default probabilities depend on the lending decision.}
	
	\textcolor{black}{The allocation problem is therefore
		\begin{align*}
			\max_{x_1,\ldots,x_N}\quad
			&\sum_{i=1}^{N}x_i\eta^i
			\left[1-\mathrm{PD}^i_T(x_i;s)\right],\\
			\text{subject to}\quad
			&\sum_{i=1}^{N}x_i=K,\\
			&0\leq x_i\leq\kappa K,\qquad i=1,\ldots,N,
		\end{align*}
		where $\kappa$ limits individual exposure concentration. The nonlinear formulation accounts for endogenous default risk, while a fixed-PD linear approximation provides a comparison benchmark. The analysis evaluates how NGFS scenario adjustments affect the resulting credit-risk and lending outcomes.}

	\section{Transition and Physical Risk Modeling}\label{transition-risk}

	\textcolor{black}{Transition risk arises from changing policies, technologies, energy use, and market demand. We describe firm emissions and production under NGFS 2024 short-term scenarios, alongside the physical-risk shocks affecting obligor creditworthiness.}
	
	\subsection{Firm-Level Emission and Cost Modeling under Climate Policy}
	
	\paragraph*{Emission Decision Vector}
	Each firm has access to a set of energy sources, denoted \(\mathcal{E}^i\). For each energy source \(e \in \mathcal{E}^i\) and each time \(t\), the firm emits \(\gamma^i_{e,t}\) kilograms of CO\(_2\) equivalent (CO\(_2\)e) per unit time. The total emission vector at time \(t\) is
	
	\[
	\gamma^i_{\bullet,t} = (\gamma^i_{e,t})_{e \in \mathcal{E}^i}, \quad \text{with total emissions } \sum_{e \in \mathcal{E}^i} \gamma^i_{e,t} = \mathbf{1} \cdot \gamma^i_{\bullet,t}.
	\]
	
	\paragraph*{Energy Output and Cost}
	
	\textcolor{black}{Let $\theta^i_{e,t}$ denote energy output per unit emission and $\alpha^i_{e,t}$ the unit energy cost. Including a quadratic term to capture nonlinear energy expenditures gives the \textit{cost function}}
	\[
	C^i(\gamma^i_{\bullet,t}) = \sum_{e \in \mathcal{E}^i} \alpha^i_{e,t} \theta^i_{e,t} \gamma^i_{e,t} + \sum_{e \in \mathcal{E}^i} \beta^i_{e,t} (\theta^i_{e,t} \gamma^i_{e,t})^2.
	\]
	Define:
	\[\begin{aligned}
		\alpha^{i,\theta}_{\bullet,t} &= (\alpha^i_{e,t} \theta^i_{e,t})_{e \in \mathcal{E}^i}, \\
		\beta^{i,\theta}_{\bullet,t} &= (\beta^i_{e,t} (\theta^i_{e,t})^2)_{e \in \mathcal{E}^i}, \\
		(\gamma^2)^i_{\bullet,t} &= ({\gamma^i_{e,t}}^2)_{e \in \mathcal{E}^i},
	\end{aligned}\]
	so that the cost function can be expressed in vector form as
	\[
	C^i(\gamma^i_{\bullet,t}) = \alpha^{i,\theta}_{\cdot,t} \cdot \gamma^i_{\bullet,t} + \beta^{i,\theta}_{\bullet,t} (\gamma^2)^i_{\bullet,t}.
	\]
	
	\paragraph*{Climate Regulation: Emission Benchmark}
	Let \(\tilde{\gamma}^i_t\) denote the benchmark (or target) emissions under a regulatory scenario. Firms face a penalty if their emissions exceed the benchmark, and a reward if their emissions are below the benchmark. 
	\textcolor{black}{Let $z_+=\max\{z,0\}$. We represent the regulatory penalty and reward by \[\ell_1^i(z)=\omega_1^i(z_+)^2, \qquad \ell_2^i(z)=\omega_2^i(z_+)^2, \]
		where the respective arguments are $z=\mathbf{1}\cdot\gamma_{\bullet,t}^i -\widetilde{\gamma}_t^i$ and $z=\widetilde{\gamma}_t^i-\mathbf{1}\cdot\gamma_{\bullet,t}^i$. Thus, excess emissions incur a penalty, whereas emissions below the benchmark generate a reward. The positive-part specification ensures that these incentives apply on their intended sides of the regulatory benchmark.}

	Together, the emission decision vector, energy cost structure, and regulatory penalty/reward scheme \textcolor{black}{describe the economic trade-offs between production, energy expenditures, emission reductions, and regulatory compliance. These considerations motivate the scenario-response specification used to incorporate climate-related shocks into firm production dynamics and credit-risk assessment.}

	\subsection{Firm Production Process}
	
	\paragraph*{Production and Emission Linkage}
	
	To model how firms respond to climate policy and lending allocation decisions, we define the evolution of a firm's logarithmic production level, denoted 
	\(p^i_t\). Log-production provides a convenient way to express percentage changes in output and ensures positivity of production over time. The stochastic differential equation governing log-production is

	\begin{equation}
		\label{eq:log_production}
		\mathrm{d}p^i_t = \mu^i(t, p^i_t, \gamma^i_{\bullet,t})\,\mathrm{d}t + \sigma^i \mathrm{d}W^i_t,
	\end{equation}
	where the drift term encodes firm-level economic and environmental dynamics
	\begin{equation}
		\label{eq:mu_modified}
		\mu^i(t, p^i_t, \gamma^i_{\bullet,t}) = a^i - b^i p^i_t + (c^{i,\theta}_\bullet \cdot \gamma^i_{\bullet,t}) - x_i \eta^i.
	\end{equation}
	Each component of this specification has a clear economic interpretation, namely,
	\begin{itemize}
		\item \(a^i\) is a baseline productivity growth rate;
		\item \(b^i > 0\) ensures mean reversion in log-production;
		\item \(c^{i,\theta}_\bullet=(c^i_e \theta^i_{e,t})_{e\in \mathcal{E}}\) captures how energy sources affect output;
		\item \(x_i\) is the amount of capital to be allocated to obligor \(i\);
		\item \textcolor{black}{\(\eta^i\) is the obligor-specific lending rate, also used as the coefficient of the exposure-related productivity adjustment in the benchmark specification.}
		\item \(\sigma^i\) is production volatility;
		\item \(W^i_t = \rho^i B_t + \sqrt{1 - (\rho^i)^2} B^i_t\) is a Brownian motion with common and idiosyncratic parts.
	\end{itemize}
	
	\textcolor{black}{The scenario-dependent drift adjustments and lending-related productivity penalty operate over the lending horizon $[0,T]$. At maturity, the continuation-value calculation uses baseline production dynamics,
		\[
		\mathrm{d}p_u^i=(a^i-b^ip_u^i)\,\mathrm{d}u
		+\sigma^i\,\mathrm{d}W_u^i,
		\qquad u>T.
		\]
		This specification allows climate scenarios and lending decisions to influence terminal creditworthiness through $p_T^i$, while retaining a common valuation rule at maturity.}

	\textcolor{black}{The term $x_i\eta^i$ represents an exposure-related adjustment to expected log-production, capturing the potential effects of debt-servicing burdens, financial stress, and reduced operational flexibility as lending exposure increases. In the benchmark specification, $\eta^i$ serves both as the lending rate and as the coefficient governing the sensitivity of production to additional exposure. This shared-parameter assumption provides a parsimonious connection between loan pricing, financial leverage, and borrower creditworthiness. Lending exposures are expressed in normalized budget units, with $K=100$, and the resulting drift adjustment is interpreted within this normalization. Emission choices influence expected production through $c_\bullet^{i,\theta}\cdot\gamma_{\bullet,t}^i$, while the Brownian component captures systemic and idiosyncratic production uncertainty.}

	\subsection{\textcolor{black}{Emission Strategies and Scenario Response}}

	\textcolor{black}{The economic motivation for a firm's emission decisions is the discounted value of future profits. For an admissible emission policy $\gamma^i$, let}
	\begin{align}
		J^i(\gamma^i):= \mathbb{E}\bigg[\int_0^\infty e^{-rt} \bigg(&AP^i e^{p_t^i}
		-C^i(\gamma_{\bullet,t}^i)\nonumber\\
		&-\ell_1^i\bigl(\mathbf{1}\cdot\gamma_{\bullet,t}^i -\widetilde{\gamma}_t^i\bigr)
		+\ell_2^i\bigl( \widetilde{\gamma}_t^i -\mathbf{1}\cdot\gamma_{\bullet,t}^i\bigr)
		\bigg)\,\mathrm{d}t\bigg].
	\end{align}
	
	\textcolor{black}{The profit functional describes the economic motivation for emission decisions, but the numerical analysis does not solve the associated stochastic control problem. Instead, we adopt a reduced-form quadratic response specification that translates NGFS scenario shocks into obligor-specific production-drift adjustments. The resulting response coefficients are signed economic transmission measures, distinct from the nonnegative physical emissions represented by $\gamma^i_{e,t}$.}

	\textcolor{black}{For each obligor $i$, climate channel $j\in\{1,2,3\}$, scenario $s$, and annual period $k$, consider the quadratic response function
		\[
		H^i_{j,s,k}(z) = \left(
		\frac{\theta^i_{j,s,k}}{r+b^i}
		-\alpha_j
		\right)z-\beta_jz^2,
		\qquad \beta_j>0.
		\]
		Because
		\[
		\frac{\mathrm{d}^2 H^i_{j,s,k}}
		{\mathrm{d}z^2}
		=-2\beta_j<0,
		\]
		this function has the unique unconstrained maximizer
		\begin{equation}
			\gamma^i_{j,s,k}
			=
			\frac{1}{2\beta_j}
			\left(
			\frac{\theta^i_{j,s,k}}{r+b^i}
			-\alpha_j
			\right).
			\label{eq:gamma_scenario_response}
		\end{equation}
		The maximization is taken over $z\in\mathbb{R}$ and defines an auxiliary scenario-response coefficient rather than an optimal emission quantity. The coefficients $\gamma^i_{j,s,k}$ may therefore take either sign, reflecting favorable or adverse contributions to expected production. Their weighted sum determines the annual climate-dependent drift adjustment, providing a direct mapping from NGFS scenario projections to firm dynamics and structural default probabilities. The numerical implementation of this transmission mechanism is detailed in the Implementation Details section.}

	\textcolor{black}{For completeness, the mean-reverting log-production dynamics are well posed for any progressively measurable drift input $\gamma^i_{\bullet,t}$ that is integrable on finite time intervals. With $b^i>0$, the unique strong solution is
		\[
		\begin{aligned}
			p_t^i={}&e^{-b^it}p_0^i\\
			&+\int_0^t e^{-b^i(t-u)}
			\left(
			a^i+
			c^{i,\theta}_{\bullet}\cdot
			\gamma^i_{\bullet,u}
			-x_i\eta^i
			\right)\mathrm{d}u\\
			&+\sigma^i\int_0^t
			e^{-b^i(t-u)}\,\mathrm{d}W_u^i.
		\end{aligned}
		\]
		This follows directly from the variation-of-constants formula for linear stochastic differential equations.}

	\subsection{Short-Term Climate Scenarios (NGFS, 2024)}

	\textcolor{black}{We map NGFS 2024 short-term sectoral shocks for 2025--2029 to firm production drift and evaluate credit risk at the beginning of 2030.}
	
	The four scenarios considered are:
	
	\begin{enumerate}
		\item \textbf{Disasters and Policy Stagnation (DAPS):} No additional climate policy, but rising physical damages from extreme weather.
		
		\item \textbf{Highway to Paris (HWTP):} An orderly transition with early, coordinated policy action and technology-driven decarbonization.
		
		\item \textbf{Sudden Wake-Up Call (SWUC):} A delayed and abrupt policy shift leading to rapid repricing and elevated transition risk.
		
		\item \textbf{Diverging Realities (DiRe):} Advanced economies transition smoothly, while other regions face escalating physical risks that spill over through trade and financial channels.

	\end{enumerate}

	\paragraph{Integration into our framework}

	\textcolor{black}{NGFS sector-year shocks modify production drift and thereby terminal default risk and lending decisions; the numerical mapping is specified in Implementation Details.}
	
	\section{Modeling the Value Process}

	\textcolor{black}{We formulate firm value as the discounted expected value of future cash flows, adjusted for climate-related losses. The resulting structural model links climate-dependent production dynamics to terminal default probabilities, providing the credit-risk measures used in the lending allocation problem.}

	\subsection{Firm Value Dynamics and Default Probability}

	\textcolor{black}{Following \cite{bourgey2025}, firm value is expected discounted profit net of physical losses:}
	\[V_t^i := \mathbb{E}\left[ \int_t^\infty e^{-r(u - t)} \left( \Pi_u^i \right)\; \textrm{d}u \;\middle|\; \mathcal{F}_t \right] - \text{EPL}^i(t),\]
	where \(\Pi_u^i\) denotes the instantaneous net profit generated by the firm at future time \(u\) conditional on the information available at time \(t\). The profit flow \(\Pi^i_u\) takes into account several components, namely 
	\begin{itemize}
		\item revenue from production \(AP^i \widehat{P}_u^i\);
		\item energy cost \(C^i_u(\widehat{{\gamma}}^i_u)\);
		\item regulatory penalties and rewards \(\ell_1^i, \ell_2^i\) based on deviation from benchmark emissions;
		\item Expected Physical Loss \(\text{EPL}^i(t)\) from climate damage.
	\end{itemize}

	\textcolor{black}{For a specified admissible emission policy $\widehat{\gamma}^i$, climate-adjusted firm value is}
	\[ \widehat{V}^i_t = \mathbb{E} \left[ \int_t^{\infty} e^{-r(u - t)} \left( AP^i \, \widehat{P}^i_u - \text{Costs}^i_u \right) \;\textrm{d}u \, \Big| \, \mathcal{F}_t \right] - \text{EPL}^i(t),\]
	where \( \widehat{P}^i_u = \exp\left( \widehat{p}^i_u \right) \), and \( \text{Costs}^i_u \) includes energy costs, penalty/reward terms, and emission adjustments (see Section~\ref{transition-risk}).
	
	Default is triggered at time \(t\) if the firm's value falls below a liability threshold \(L^i(t)\). Accordingly, the probability of default is defined as
	\[ \text{PD}^i_t := \mathbb{P}\left( \widehat{V}^i_t \leq L^i(t) \right).\]
	
	The log-production process \( \widehat{p}^i_u \) follows a linear SDE \textcolor{black}{incorporating the effects of emissions and lending exposure}	
	\[ \textrm{d}\widehat{p}^i_u = \left( a^i - b^i \widehat{p}^i_u + c_{\bullet} \cdot \widehat{\gamma}^i_u - x_i \eta^i \right)\; \textrm{d}u + \sigma^i \; \textrm{d}W^i_u,\]
	\textcolor{black}{where $x_i$ is the loan allocated to obligor $i$ and $\eta^i$ determines the corresponding leverage-related drift adjustment} (cf. Eq.~\eqref{eq:mu_modified}). Solving this SDE gives
	\[\widehat{p}^i_u = e^{-b^i(u-t)} \widehat{p}^i_t + \dfrac{a^i}{b^i}\left(1-e^{-b^i(u-t)}\right) + \displaystyle \int_{t}^{u} e^{-b^i(u-s)} \left( c_{\bullet}^{i,\theta} \widehat{\gamma}_{\bullet,s}^i-x_i\eta^i\right)\; \textrm{d}s+\sigma^i \int_t^u e^{-b^i(u-s)} \; \textrm{d}W^i_s,\]
	for \textcolor{black}{\(0\leq t\leq u\leq T\)}, where 
	\(\textrm{d}W^i_s = \rho^i \;\textrm{d}B_s + \sqrt{1-(\rho^i)^2} \; \textrm{d}B^i_s\),
	and with the physical-risk-adjusted mean
	\begin{equation} \label{eq:mu_tilde}
		\widetilde{\mu}^i(u, t) := \frac{a^i}{b^i} \left(1 - e^{-b^i(u - t)}\right)
		+ \int_t^u e^{-b^i(u - s)} \left( c_{\bullet} \cdot \widehat{\gamma}^i_s \right) ds
		- \frac{x_i \eta^i}{b^i} \left(1 - e^{-b^i(u - t)}\right).
	\end{equation}

	\textcolor{black}{Under the NGFS scenario-response specification, the climate-adjusted drift inputs are deterministic conditional on the selected scenario and obligor characteristics, with piecewise-constant annual values over the scenario horizon. Consequently, conditional on $\mathcal{F}_t$, the vector $(\widehat{p}_u^i)_{1\leq i\leq N}$ is Gaussian, with conditional mean}	
	\begin{equation}
		\mathbb{E}[\widehat{p}^i_u\mid\mathcal{F}_t] \quad = e^{-b^i(u-t)}\widehat{p}^i_t + \widetilde{\mu}^i(u,t)
	\end{equation}
	and conditional variance is given as 
	\begin{equation}
		\textrm{Cov} \left(\widehat{p}^i_u, \widehat{p}^j_u \mid \mathcal{F}_t\right) := \Sigma^{i,j}(u-t) = \sigma^i\sigma^j \left(\rho^i\rho^j \mathbf{1}_{i\neq j} + \mathbf{1}_{i=j}\right) \dfrac{1-e^{-(b^i+b^j)(u-t)}}{b^i+b^j},
	\end{equation}
	for every \(i,j\in \{1,\ldots, n\}\).

	\subsection{Firm Value as a Function of Log-Production}

	\textcolor{black}{To evaluate the default condition, we express firm value as a function of the current log-production level. Under the deterministic scenario-response specification, future log-production is conditionally Gaussian, and the corresponding production level $\widehat{P}_u^i=\exp(\widehat{p}_u^i)$ is conditionally lognormal. Its conditional expectation is therefore}
	
	\[\mathbb{E}\left[\widehat{P}^i_u\mid \mathcal{F}_t\right] = e^{e^{-b^i(u-t)}\widehat{p}^i_t + \widetilde{\mu}^i(u,t) + \frac{1}{2}\Sigma^{i,i}(u-t)}.\]
	This allows us to compute the conditional expectation of future production in closed form and derive a deterministic mapping from the current state $\widehat{p}^i_t$ to the firm’s value $\widehat{V}^i_t$. Specifically, we define a function $h^i(t, x)$ that expresses the firm value as a deterministic function of time and the current log-production level,
	$$ \widehat{V}^i_t = h^i(t, \widehat{p}^i_t),$$
	where the deterministic function \(h^i(-,-)\) is defined for all \(t\geq 0\) and \(x\in \mathbb{R}\) as 
	\begin{equation}\label{eq:h_definition}
		\begin{aligned} 
			h^i(t,x):= {} &\textrm{AP}^i \int_t^\infty e^{-r(u-t)}
			\exp\left(e^{-b^i(u-t)}x +\widetilde{\mu}^i(u,t) +\frac{1}{2}\Sigma^{i,i}(u-t)
			\right)\,\mathrm{d}u \\
			- & \int_{t}^\infty e^{-r(u-t)} \left( \alpha_{\bullet,u}^{i,\theta} \widehat{\gamma}_{\bullet,u}^i + \beta_{\bullet,u}^{i,\theta} (\widehat{\gamma}^2)_{\bullet,u}^i + \omega_1^i(\mathbf{1}\cdot \widehat{\gamma}_{\bullet,u}^i - \tilde{\gamma}^i_u)^2_+ - \omega_2^i (\tilde{\gamma}^i_u - \mathbf{1}\cdot \widehat{\gamma}_{\bullet,u}^i)^2_+\right) \; \textrm{d} u \\
			- & \textrm{EPL}^i(t).
		\end{aligned} 
	\end{equation} 
	This function encodes the effects of stochastic production growth, emission-dependent costs, climate damages, and discounting, and serves as a key input in computing default probabilities under our structural framework. Notably, for each fixed $t$, the mapping $x \mapsto h^i(t, x)$ is strictly increasing, continuous, and invertible.

	\subsection{The Auxiliary Process and Default Probability}

	Now, since \(\widehat{V}^i_t =h^i(t, \widehat{p}^i_t)\), we have that 
	\[\begin{aligned}
		\mathbb{P}\left(\widehat{V}^i_t \leq L^i(t)\right) &= \mathbb{P}\left(h^i(t,\widehat{p}^i_t)\leq L^i(t)\right)\\
		&= \mathbb{P}\left( \widehat{p}^i_t \leq h^i(t,-)^{-1}(L^i(t))\right) \\
		&=\mathbb{P} \Bigg( e^{-b^it} \widehat{p}^i_0 +\widetilde{\mu}^i(t,0) + \sigma^i\rho^i \int_0^t e^{-b^i(t-s)}\;\textrm{d}B_s\\ 
		&+\sigma^i \sqrt{1-(\rho^i)^2} \int_0^t e^{-b^i (t-s)}\; \textrm{d}B^i_s \leq h^i(t,-)^{-1}(L^i(t)) \Bigg)\\
		& = \mathbb{P} \Bigg( e^{-b^it} \widehat{p}^i_0 +\widetilde{\mu}^i(t,0) +\sigma^i \sqrt{1-(\rho^i)^2} \int_0^t e^{-b^i (t-s)}\; \textrm{d}B^i_s  \\ 
		& - h^i(t,-)^{-1}(L^i(t))  \leq - \sigma^i\rho^i \int_0^t e^{-b^i(t-s)}\;\textrm{d}B_s \Bigg)\\
		&= \mathbb{P} (A_t^i\leq X_t^i),
	\end{aligned}\]
	where 
	\begin{eqnarray*}\label{A_definition}
		A^i_t &:=& \dfrac{1}{\sigma^i} \left( e^{-b^it}\widehat{p}^i_0 +\widetilde{\mu}^i(t,0) + \sigma^i \sqrt{1-(\rho^i)^2} \int_0^t e^{-b^i(t-s)}\; \textrm{d}B^i_s - h^i(t,-)^{-1}(L^i(t)) \right) \\
		X^i_t&:=& - \rho^i \int_0^t e^{-b^i(t-s)}\; \textrm{d}B_s. 
	\end{eqnarray*}

	\textcolor{black}{\subsection{Principal Component Representation of Systemic Risk}}
	
	\textcolor{black}{Dependence across obligors is introduced through a common Brownian factor in the log-production dynamics. Let $X_T=(X_T^1,\ldots,X_T^N)^\top$ denote the resulting systemic component at the terminal horizon. Its covariance matrix has entries
		\[(K_{X_T})^{i,j}=\rho^i\rho^j \frac{1-e^{-(b^i+b^j)T}}{b^i+b^j}.\]
		Since $K_{X_T}$ is positive semidefinite, its spectral decomposition provides a Gaussian factor representation. Retaining the leading $q$ eigenpairs yields
		\[X_T\approx \sum_{k=1}^{q}\sqrt{\nu^k}\,G^ku^k, \qquad G^k\overset{\mathrm{iid}}{\sim}\mathcal{N}(0,1),\]
		where $\nu^k$ and $u^k$ denote the eigenvalues and corresponding eigenvectors of $K_{X_T}$.}

	\textcolor{black}{In the benchmark implementation, $q=5$ retains more than $99.9999\%$ of the model-implied systemic covariance variance. Let
		\[K_{X_T}^{(q)}=\sum_{k=1}^{q}\nu^k u^k(u^k)^\top\]
		denote the retained covariance approximation, and define the discarded marginal systemic variance by
		\[\delta_i=(K_{X_T})^{i,i}-(K_{X_T}^{(q)})^{i,i}\geq 0.\]
		Writing
		\[q_2^i=\frac{1-e^{-2b^iT}}{2b^i},\]
		the standardized Gaussian latent variable used in the joint-default simulation is
		\[Y_i=\frac{\displaystyle\sum_{k=1}^{q}\sqrt{\nu^k}\,u_i^kG^k+ \sqrt{(1-(\rho^i)^2)q_2^i+\delta_i}\,Z_i}{\sqrt{q_2^i}},\]
		where $G^1,\ldots,G^q,Z_1,\ldots,Z_N$ are independent standard normal variables.}
	
	\textcolor{black}{The independent residual term preserves unit marginal variance and hence each analytical marginal PD, while the retained components approximate cross-obligor covariance. Other component counts and tail dependence are tested below.}

	\subsection{Monte Carlo Simulation}

	\textcolor{black}{Gaussian marginal default probabilities are analytical; Monte Carlo draws are used for joint losses and numerical validation.}
	
	\textcolor{black}{For each scenario, we generate $M=6000$ realizations of the systemic and idiosyncratic components. Systemic dependence is represented through the retained principal components of the model-implied covariance matrix, with residual variance included to preserve marginal distributions. For obligor $i$, the simulated default probability is
		\[\widehat{\mathrm{PD}}^i_T=\frac{1}{M}\sum_{m=1}^{M} \mathbbm{1}_{\{Y_i^{(m)} \leq \Phi^{-1}(\mathrm{PD}^i_T)\}}, \] where $Y_i^{(m)}$ is the standardized Gaussian latent variable and $\Phi$ denotes the standard normal distribution function.}
	
	\textcolor{black}{Common simulated latent states enable paired policy comparisons.}

	\textcolor{black}{\subsection{Climate-Aware Lending Optimization}}
	
	\textcolor{black}{We consider a financial institution allocating a fixed lending budget $K$ across $N$ obligors under climate scenario $s$. The allocation to obligor $i$ is denoted by $x_i$, and $\eta^i$ represents its lending rate. Because lending affects firm performance through the leverage-related drift penalty, the terminal default probability $\mathrm{PD}_T^i(x_i;s)$ depends on the allocated amount.}
	
	\textcolor{black}{The lender maximizes expected surviving-interest income:
		\begin{align}
			\max_{x_1,\ldots,x_N}\quad
			&\sum_{i=1}^{N}x_i\eta^i
			\left[1-\mathrm{PD}_T^i(x_i;s)\right],
			\label{eq:nonlinear_allocation}\\
			\text{subject to}\quad
			&\sum_{i=1}^{N}x_i=K,\nonumber\\
			&0\leq x_i\leq\kappa K,\qquad i=1,\ldots,N.\nonumber
		\end{align}
		Here, $\kappa$ limits individual exposure concentration. Unlike a conventional linear allocation problem, the objective is nonlinear because default probabilities respond endogenously to lending decisions.}

	\textcolor{black}{We solve the constrained problem numerically using sequential quadratic programming (SLSQP), with $K=100$ and a benchmark diversification limit of $\kappa=0.05$. The algorithm provides a locally optimal solution satisfying the numerical convergence and feasibility criteria; global optimality is not assumed.}
	
	\textcolor{black}{To assess the contribution of climate information, we compare scenario-specific allocations with a conventional allocation obtained without climate adjustments and an equal-weight portfolio. We also consider a fixed-PD linear approximation in which default probabilities are evaluated at the reference allocation $x_i=K/N$ and subsequently held constant during optimization. All resulting portfolios are evaluated using default probabilities recalculated at their actual allocations, ensuring a consistent comparison.}
	
	\textcolor{black}{Expected credit losses are reported separately using a benchmark loss-given-default assumption of $45\%$. They provide a complementary measure of portfolio risk but do not enter the surviving-interest objective directly.}

	\section{Implementation Details}

	\textcolor{black}{The R implementation uses four NGFS scenarios with annual 2025--2029 inputs: GEME3 sector production, CLIMACRED financial adjustments, and available DAPS/DiRe physical impacts. These inputs are harmonized over 50 sectors for a synthetic North American loan portfolio.}
	
	\textcolor{black}{A common population of 1,000 synthetic obligors permits within-portfolio scenario comparisons. Institution-specific exposures, ratings, financial characteristics, spreads, and defaults would be needed for empirical calibration.}

	\paragraph*{Portfolio Construction and Calibration}
	
	\textcolor{black}{Obligors are assigned to 50 sectors with weights proportional to the square root of baseline GEME3 production, rather than assumed bank lending shares. Assignments and borrower parameters remain fixed across scenarios.}
	
	\textcolor{black}{Firm-level heterogeneity is introduced through
		\[p_0^i=0,\qquad a^i\sim\mathcal{U}(0,0.5),\qquad b^i\sim\mathcal{U}(1,4),\]
		\[\sigma^i\sim\mathcal{U}(0.1,0.5),\qquad \rho^i\sim\mathcal{U}(-1,1).\]
		These bounded distributions represent variation in baseline productivity, mean reversion, production volatility, and systemic exposure. They provide a transparent benchmark calibration for evaluating structural and scenario sensitivities. Lending rates are specified as
		\[\eta^i=r+\Lambda_i,\qquad r=0.02,\qquad \Lambda_i\sim\mathcal{U}(0.025,0.06),\]
		replacing the previous specification in which lending rates depended on the obligor index.}
	
	\textcolor{black}{The structural default barrier is calibrated to a reference annual hazard rate of $\lambda_{\mathrm{ref}}=0.03$, corresponding to a five-year terminal default probability of $1-e^{-0.15}\approx 13.93\%$. This provides a common reference level rather than an estimate of the average default risk of an observed banking portfolio. A loss-given-default assumption of $45\%$ is used to calculate expected credit losses for comparison purposes. The small reference valuation adjustment, set at $10^{-5}$ of firm value, is a numerical convention and should not be interpreted as an empirically calibrated physical-damage rate; scenario-dependent physical effects are incorporated separately through the NGFS inputs.}
	
	\textcolor{black}{The climate-to-drift mapping retains fixed sensitivity coefficients, with $c_j=0.01$, $\alpha_j=0$, and $\beta_j\in\{0.1,0.5,0.8\}$. These coefficients specify the benchmark transmission intensity rather than estimated causal effects. Their implications are examined through alternative shock mappings and intensity assumptions.}
	
	\textcolor{black}{Systemic dependence is represented using the leading five eigenvectors of the model-implied covariance matrix, retaining more than $99.9999\%$ of its variance. The lending budget is normalized to $K=100$, with a benchmark individual-exposure limit of $\kappa=0.05$. Unlike the previous linear-programming implementation, this limit is nonbinding in the principal nonlinear solution; tighter limits are therefore examined to assess concentration sensitivity. Additional robustness experiments vary structural parameters, PCA dimensions, and dependence assumptions.}

\paragraph*{Climate Scenario Integration and Drift Adjustment}

\textcolor{black}{The five annual NGFS observations are treated as piecewise-constant drift shocks on $[k-1,k)$, $k=1,\ldots,5$, with terminal risk evaluated at the beginning of 2030.}

	\textcolor{black}{For obligor $i$, scenario $s$, and annual period $k$, let $\theta^i_{j,s,k}$ denote the shock associated with channel $j\in\{1,2,3\}$. The production channel is the deviation of scenario production from baseline production, with the underlying production ratio bounded between $0.25$ and $4$ as a numerical safeguard. The financial channel is the negative of the CLIMACRED corporate bond-spread adjustment, while the physical channel is the negative sum of productivity loss, production loss, and capital destruction. The direct physical-impact channel is included for DAPS and DiRe, for which the corresponding source data are available.}
	
	\textcolor{black}{Following the structural specification, the climate contribution to the annual production drift is}
	\[d^i_{s,k} = \sum_{j=1}^{3}c_j\gamma^i_{j,s,k}, \qquad \gamma^i_{j,s,k} =\frac{1}{2\beta_j}
	\left(\frac{\theta^i_{j,s,k}}{r+b^i}-\alpha_j \right).\]

	\textcolor{black}{Because mean reversion offsets older shocks, an adverse shock in 2025 has less influence on terminal creditworthiness than the same shock in 2029. Define
		\[	w^i_k = \frac{e^{-b^i(T-k)} - e^{-b^i(T-k+1)}}{b^i}, \qquad k=1,\ldots,5, \]
		so that
		\[ 	S^i_s=\sum_{k=1}^{5}w^i_k d^i_{s,k}.\]}
	
\paragraph*{Reference Default Calibration}
	
	\textcolor{black}{Let
		\[q_1^i=\frac{1-e^{-b^iT}}{b^i}, \qquad q_2^i=\frac{1-e^{-2b^iT}}{2b^i},\]
		and define the reference allocation $x_i^{\mathrm{ref}}=K/N$.}
	
	\textcolor{black}{Without climate adjustments, the terminal log-production mean at this allocation is
		\[m_i^{\mathrm{ref}}=e^{-b^iT}p_0^i	+a^iq_1^i -\eta^ix_i^{\mathrm{ref}}q_1^i.\]
		The reference terminal default probability is
		\[p_{\mathrm{ref}}= 1-e^{-\lambda_{\mathrm{ref}}T}, \qquad \lambda_{\mathrm{ref}}=0.03.\]
		Accordingly, the reference log-production default threshold is calibrated as
		\[z_i=m_i^{\mathrm{ref}}+ \sigma^i\sqrt{q_2^i}\, \Phi^{-1}(p_{\mathrm{ref}}), \]
		where $\Phi$ is the standard normal distribution function.}
	
	\textcolor{black}{For a candidate allocation $x_i$ under scenario $s$, terminal log-production has conditional mean
		\[m_i(x_i;s)= e^{-b^iT}p_0^i+a^iq_1^i +S_s^i -\eta^ix_iq_1^i\]
		and variance $(\sigma^i)^2q_2^i$. Therefore,
		\[\mathrm{PD}_T^i(x_i;s)= \Phi\left( \frac{z_i-m_i(x_i;s)}{\sigma^i\sqrt{q_2^i}} \right).\]
		The reference threshold is held fixed across climate scenarios and lending policies, so differences in default probabilities reflect scenario shocks and changes in lending exposure rather than repeated recalibration of the default barrier.}

	\textcolor{black}{The calibrated log-production threshold $z_i$ is linked to a structural liability threshold through a common baseline continuation-value function at maturity. For $v\geq 0$, define
		\[
		q_1^i(v)=\frac{1-e^{-b^iv}}{b^i},
		\qquad
		q_2^i(v)=\frac{1-e^{-2b^iv}}{2b^i}.
		\]
		The benchmark terminal continuation-value mapping is
		\[
		\begin{aligned}
			h_0^i(T,p)
			={}&AP^i\int_0^\infty e^{-rv}
			\exp\left(
			e^{-b^iv}p
			+a^iq_1^i(v)
			+\frac{(\sigma^i)^2}{2}q_2^i(v)
			\right)\,\mathrm{d}v\\
			&-\mathrm{EPL}_{\mathrm{ref}}^i.
		\end{aligned}
		\]
		Here $\mathrm{EPL}_{\mathrm{ref}}^i$ is a fixed reference valuation adjustment. The benchmark continuation valuation uses post-maturity baseline production cash flows; scenario-specific emission and regulatory costs are transmitted through the production dynamics over the lending horizon and are not separately activated in this terminal continuation calculation.}
	
	\textcolor{black}{The corresponding fixed liability threshold is defined by
		\[
		L^i(T)=h_0^i(T,z_i).
		\]
		Because $AP^i>0$ and $b^i>0$, the mapping $p\mapsto h_0^i(T,p)$ is continuous and strictly increasing. Consequently,
		\[
		\begin{aligned}
			\mathrm{PD}_T^i(x_i;s)
			&=\mathbb{P}\left(
			h_0^i(T,p_T^i(x_i;s))
			\leq L^i(T)
			\right)\\
			&=\mathbb{P}\left(p_T^i(x_i;s)\leq z_i\right)\\
			&=\Phi\left(
			\frac{z_i-m_i(x_i;s)}
			{\sigma^i\sqrt{q_2^i}}
			\right).
		\end{aligned}
		\]
		Thus, the structural liability threshold and its log-production counterpart remain fixed across scenarios and lending policies. The numerical implementation uses analytical Gaussian default probabilities and evaluates the continuation-value mapping by quadrature for representative obligors to check its monotonicity.}

\paragraph*{Lending Allocation and Numerical Solution}
	
	\textcolor{black}{For each scenario, \eqref{eq:nonlinear_allocation} is solved using R \texttt{nloptr::slsqp}, with analytical objective gradients and exposure-dependent PDs. Budget feasibility, exposure bounds, and numerical convergence are checked.}
	
	\textcolor{black}{Conventional allocation solves the same nonlinear problem without climate shocks; equal-weight and fixed-PD linear allocations provide further comparisons. All policies are assessed using PDs recalculated at their actual exposures.}
	
\paragraph*{Joint Default Simulation and Portfolio Losses}
	
	\textcolor{black}{Joint losses use $M=6000$ draws and five systemic principal components. Independent residuals preserve the discarded marginal variance and idiosyncratic risk.}
	
	\textcolor{black}{For simulated latent variable $Y_i^{(m)}$, the terminal default indicator is
		\[D_i^{(m)}(x_i;s)= \mathbbm{1}_{\left\{Y_i^{(m)}\leq \Phi^{-1}\bigl(\mathrm{PD}_T^i(x_i;s)\bigr)\right\}}.\]
		The portfolio loss in simulation $m$ is
		\[\mathcal{L}^{(m)}(x;s)=\mathrm{LGD} \sum_{i=1}^{N}x_iD_i^{(m)}(x_i;s), \qquad
		\mathrm{LGD}=0.45.\]
		Expected credit losses can also be computed analytically as
		\[\mathrm{ECL}(x;s)= 0.45\sum_{i=1}^{N} x_i\mathrm{PD}_T^i(x_i;s).\]}

	\textcolor{black}{Common latent draws support paired comparisons. The baseline uses Gaussian dependence; robustness employs a Student-$t$ copula with $\nu=4$.}
	
	\textcolor{black}{Let $Y_i^{(G,m)}$ denote the standardized Gaussian latent variable for obligor $i$ in simulation $m$. For each simulation, draw a common scaling variable
		\[W^{(m)}\sim\chi^2_{\nu},\]
		independently of the Gaussian latent vector, and define
		\[T_i^{(m)}=\frac{Y_i^{(G,m)}}{\sqrt{W^{(m)}/\nu}}.\]
		Using the cumulative distribution function $F_{t_\nu}$ of the standard Student-$t$ distribution, construct
		\[U_i^{(m)}=F_{t_\nu}\bigl(T_i^{(m)}\bigr), \qquad Y_i^{(t,m)}=\Phi^{-1}\bigl(U_i^{(m)}\bigr).\]
		The transformed variable $Y_i^{(t,m)}$ has a standard normal marginal distribution, while the common scaling variable generates Student-$t$ copula dependence across obligors. The corresponding default indicators are
		\[D_i^{(t,m)}(x_i;s)= \mathbbm{1}_{\left\{Y_i^{(t,m)} \leq \Phi^{-1}\bigl(\mathrm{PD}_T^i(x_i;s)\bigr)\right\}}.\]
		Thus, the analytical marginal default probabilities are preserved under both dependence specifications, allowing differences in simulated portfolio tail losses to be attributed to the dependence structure.}

	\section{Results}

	\textcolor{black}{We first compare terminal default risk at reference exposures, then examine sectoral reallocations and the economic performance of alternative lending policies.}

	\textcolor{black}{\subsection{Scenario-Dependent Credit Risk}}
	
	\textcolor{black}{Table~\ref{tab:scenario_pd_distribution} summarizes the distribution of five-year terminal default probabilities under the four NGFS scenarios. The calculations use the same 1,000 synthetic obligors, sector assignments, and structural parameters, with the equal-weight reference allocation $x_i=K/N$ held fixed across scenarios. The reference five-year probability of default without climate adjustments is approximately $13.929\%$.}
	
	\begin{table}[htbp]
		\centering
		\footnotesize
		\setlength{\tabcolsep}{4pt}
		\caption{Distribution of five-year terminal default
			probabilities across obligors (\%)}
		\label{tab:scenario_pd_distribution}
		\begin{tabular}{lrrrrrrr}
			\hline
			Scenario & Mean & Median & SD &
			$P_5$ & $P_{95}$ & Min. & Max. \\
			\hline
			DAPS & 13.955 & 13.944 & 0.030 &
			13.932 & 14.019 & 13.930 & 14.128 \\
			HWTP & 13.878 & 13.939 & 0.560 &
			13.649 & 14.190 & 6.273 & 15.930 \\
			SWUC & 13.892 & 13.945 & 0.552 &
			13.636 & 14.215 & 6.315 & 15.917 \\
			DiRe & 13.899 & 13.948 & 0.591 &
			13.640 & 14.250 & 6.164 & 15.745 \\
			\hline
		\end{tabular}
		
		\smallskip
		\parbox{0.95\textwidth}{\footnotesize
			\textit{Notes:} Statistics are calculated across
			1,000 obligors at the reference allocation.
			SD denotes cross-sectional standard deviation;
			$P_5$ and $P_{95}$ denote the fifth and
			ninety-fifth percentiles.}
	\end{table}
	
	\textcolor{black}{Mean default probabilities remain close to the reference level, ranging from $13.878\%$ under HWTP to $13.955\%$ under DAPS. The difference between these scenarios is approximately $7.7$ basis points. However, the relatively small variation in portfolio-wide averages conceals substantial differences in the cross-sectional distribution of default risk. Under DAPS, the standard deviation of default probabilities is approximately $0.03$ percentage points, compared with $0.55$--$0.59$ percentage points under HWTP, SWUC, and DiRe.}
	
	\textcolor{black}{HWTP exhibits a wider PD range (6.27--15.93\%) than DAPS, although its central 90\% lies between 13.65\% and 14.19\%. Climate effects are therefore concentrated in selected exposures.}
	
\begin{figure}[htbp]
		\centering
		\includegraphics[width=\textwidth]
		{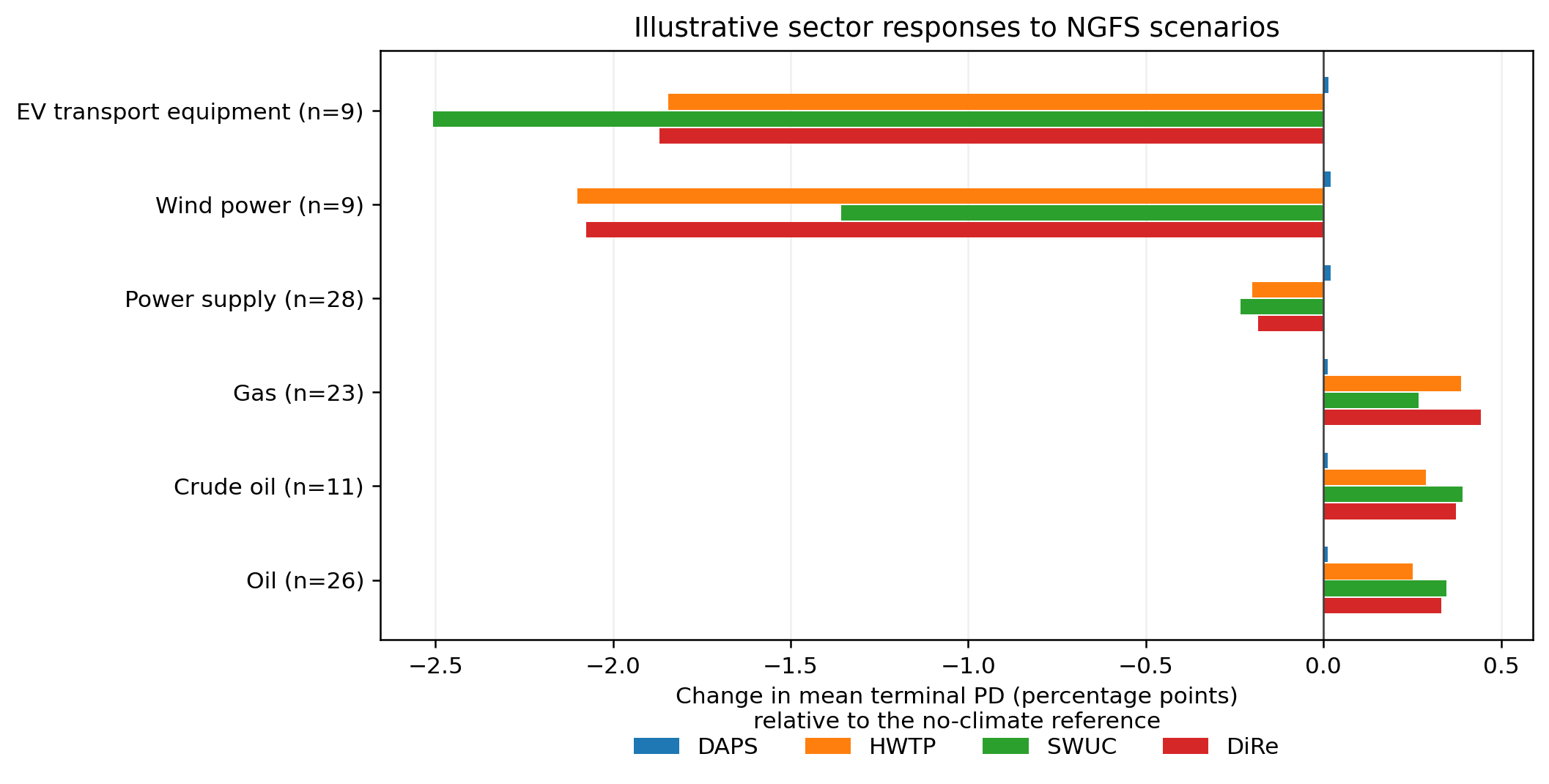}
		\caption{Changes in sector-level mean terminal default probabilities relative to the no-climate reference. Selected sectors illustrate differences in model-implied credit-risk responses across NGFS pathways; $n$ denotes the number of synthetic 	obligors represented in each sector.}
		\label{fig:sector_pd_changes}
	\end{figure}

	\textcolor{black}{Figure~\ref{fig:sector_pd_changes} illustrates these differences for selected sectors. Under HWTP, the mean default probability among wind power obligors falls to approximately $11.83\%$, compared with the reference level of $13.93\%$. Electric-vehicle transport equipment exhibits a similar response, with a mean default probability of approximately $11.42\%$ under SWUC. Conversely, the mean default probability for gas-sector obligors rises to approximately $14.32\%$ under HWTP and $14.37\%$ under DiRe. Oil and crude-oil exposures exhibit smaller increases across these scenarios.}
	
	\textcolor{black}{These results reveal an important distinction between aggregate credit-risk effects and the redistribution of risk across sectors. Scenario shocks can simultaneously improve the creditworthiness of some obligors and weaken that of others, producing substantial sectoral adjustments even when the average default probability changes only modestly. The resulting heterogeneity provides an economic rationale for evaluating climate-aware lending decisions at the obligor and sector levels rather than relying exclusively on portfolio-wide default statistics.}

	\subsection{\textcolor{black}{Scenario-Dependent Lending Reallocation}}
	
	\textcolor{black}{Climate-related changes in creditworthiness affect lending decisions through the endogenous relationship between loan exposure and default risk. The scenario-specific nonlinear allocation problem is solved under the common budget $K=100$ and benchmark individual-exposure limit $\kappa=0.05$.}
	
\begin{figure}[htbp]
		\centering
		\includegraphics[width=\textwidth]
		{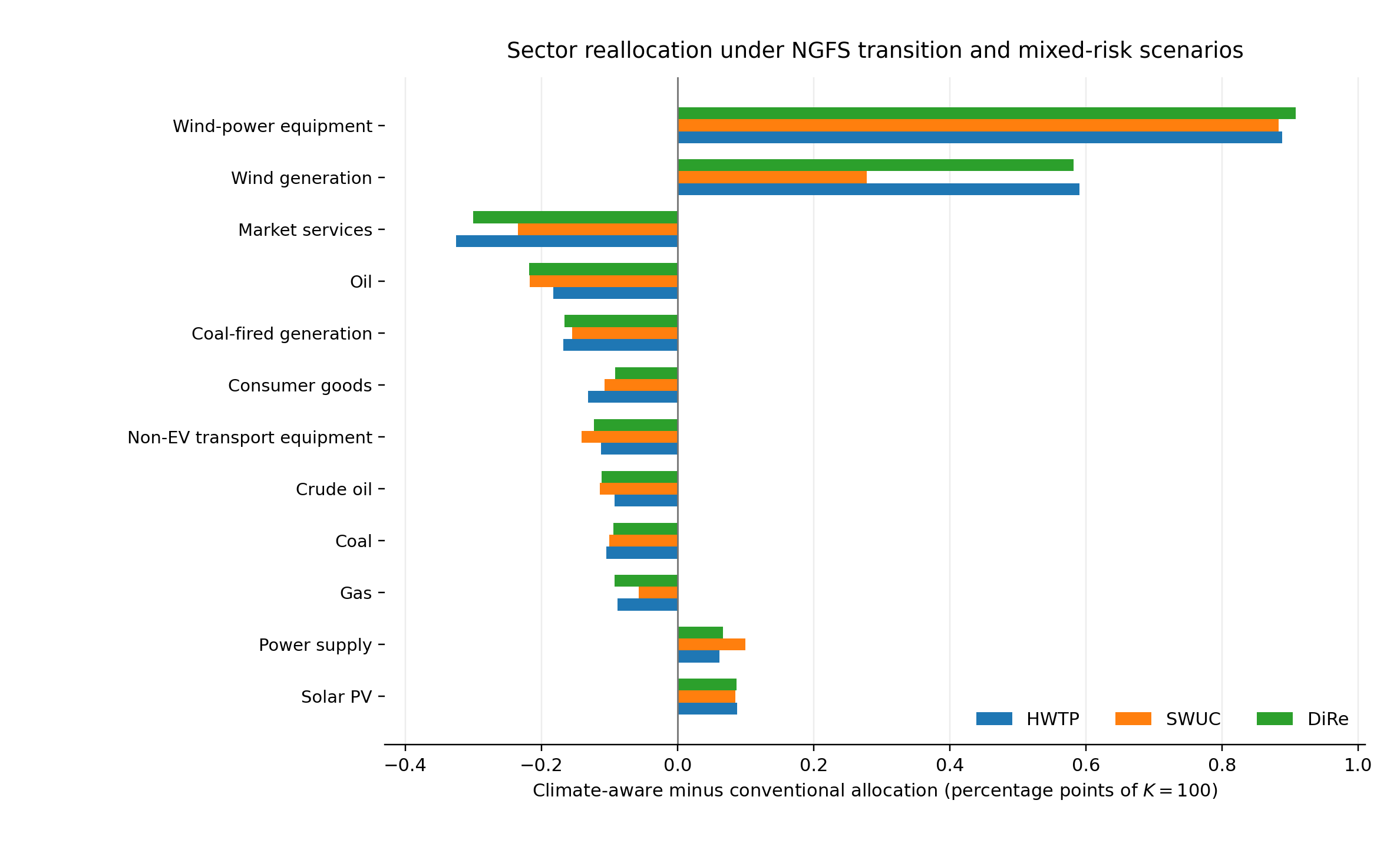}
		\caption{Changes in sectoral lending allocations relative to the conventional policy. Positive values indicate increased lending under climate-aware optimization. Changes are expressed in percentage points of the normalized lending budget $K=100$. DAPS is omitted from the figure because its allocation changes are comparatively small.}
		\label{fig:sector_allocation_shifts}
	\end{figure}

	\textcolor{black}{Figure~\ref{fig:sector_allocation_shifts} reports changes in sectoral allocations relative to conventional lending. Under HWTP, approximately 1.82 units of the total lending budget are redistributed across sectors. The corresponding amounts are 1.57 under SWUC and 1.85 under DiRe, compared with only 0.08 under DAPS. These figures represent the sum of positive sectoral allocation changes, which equals one-half of the sum of their absolute changes because the lending budget is fixed.}
	
	\textcolor{black}{The largest increases occur in wind-related activities. Under HWTP, lending to wind-power equipment increases by approximately 0.89 budget units, while the allocation to wind generation rises from 0.48 to 1.07. Similar increases in wind-power equipment occur under SWUC and DiRe. These adjustments are accompanied by reductions in several fossil-fuel-related sectors. For example, oil-sector lending declines by approximately 0.18 units under HWTP and 0.22 units under both SWUC and DiRe. Coal-fired generation and conventional transport-equipment sectors also experience reductions.}
	
	\textcolor{black}{The reallocations are concentrated in a relatively small set of sectors, consistent with the heterogeneity in default probabilities reported above. At the obligor level, the nonlinear solution funds approximately 237--238 borrowers out of 1,000. The largest individual exposure represents approximately 1.26--1.27\% of the lending budget, leaving the benchmark 5\% exposure cap nonbinding. Thus, sectoral adjustments arise primarily from changes in relative creditworthiness and the endogenous lending objective rather than an imposed tightening of diversification constraints.}
	
	\subsection{\textcolor{black}{Economic Implications of Lending Allocation}}
	
	\textcolor{black}{We evaluate four lending policies under each NGFS scenario: conventional nonlinear allocation without climate adjustments, scenario-specific climate-aware nonlinear allocation, equal-weight allocation, and a fixed-PD linear approximation. Each policy is evaluated using default probabilities recalculated at its actual loan exposures. Table~\ref{tab:policy_comparison} presents the expected surviving-interest objective, expected credit losses, and portfolio concentration measures.}
	
	\begin{table}[htbp]
		\centering
		\footnotesize
		\setlength{\tabcolsep}{4pt}
		\caption{Economic performance of alternative lending policies}
		\label{tab:policy_comparison}
		\begin{tabular}{llrrrrr}
			\hline
			Scenario & Policy & Objective & ECL & $N_+$ & Max. (\%) & HHI \\
			\hline
			DAPS & Conventional & 6.4644 & 7.3013 & 238 & 1.26 & 0.00612 \\
			& Climate-aware & 6.4644 & 7.3009 & 237 & 1.26 & 0.00613 \\
			& Equal-weight & 5.3945 & 6.2795 & 1000 & 0.10 & 0.00100 \\
			& Fixed-PD LP & 3.7787 & 23.6498 & 20 & 5.00 & 0.05000 \\
			\hline
			HWTP & Conventional & 6.4652 & 7.2967 & 238 & 1.26 & 0.00612 \\
			& Climate-aware & 6.4686 & 7.2459 & 238 & 1.27 & 0.00610 \\
			& Equal-weight & 5.3992 & 6.2449 & 1000 & 0.10 & 0.00100 \\
			& Fixed-PD LP & 3.9019 & 22.9474 & 20 & 5.00 & 0.05000 \\
			\hline
			SWUC & Conventional & 6.4642 & 7.3026 & 238 & 1.26 & 0.00612 \\
			& Climate-aware & 6.4672 & 7.2628 & 238 & 1.27 & 0.00612 \\
			& Equal-weight & 5.3982 & 6.2515 & 1000 & 0.10 & 0.00100 \\
			& Fixed-PD LP & 3.6718 & 24.2539 & 20 & 5.00 & 0.05000 \\
			\hline
			DiRe & Conventional & 6.4640 & 7.3038 & 238 & 1.26 & 0.00612 \\
			& Climate-aware & 6.4676 & 7.2518 & 238 & 1.27 & 0.00611 \\
			& Equal-weight & 5.3979 & 6.2544 & 1000 & 0.10 & 0.00100 \\
			& Fixed-PD LP & 3.8993 & 22.9617 & 20 & 5.00 & 0.05000 \\
			\hline
		\end{tabular}
		
		\smallskip
		\parbox{0.96\textwidth}{\footnotesize
			\textit{Notes:} Objective denotes expected surviving-interest income. ECL denotes analytical expected credit loss under a 45\% LGD. Both are expressed in units of the normalized lending budget. $N_+$ is the number of funded obligors; Max. is the largest individual exposure as a percentage of the budget; and HHI is $\sum_i(x_i/K)^2$. All policies are evaluated with exposure-dependent default probabilities.}
	\end{table}
	
	\textcolor{black}{The comparison reveals a trade-off between expected interest income, credit losses, and diversification. Under HWTP, climate-aware allocation achieves an expected surviving-interest objective of 6.4686, compared with 6.4652 under conventional allocation and 5.3992 under equal-weight lending. The equal-weight portfolio achieves lower expected credit losses, reflecting its broader diversification and lower exposure-weighted default risk. In contrast, the nonlinear optimized portfolios concentrate lending among borrowers offering higher expected surviving-interest contributions.}
	
	\textcolor{black}{The fixed-PD linear approximation illustrates the importance of accounting for the feedback between loan size and credit risk. With default probabilities held constant during allocation, the linear solution assigns the entire budget to 20 borrowers at the 5\% exposure limit. Recalculating default probabilities at these exposures produces expected credit losses of approximately 22.95--24.25 across scenarios, compared with 7.25--7.30 under climate-aware nonlinear allocation. The fixed-PD approximation consequently performs poorly when evaluated using the endogenous structural model.}
	
	\textcolor{black}{Table~\ref{tab:economic_comparison} reports changes relative to conventional allocation.}
	
	\begin{table}[htbp]
		\centering
		\footnotesize
		\setlength{\tabcolsep}{5pt}
		\caption{Incremental effects of climate-aware lending relative to conventional allocation}
		\label{tab:economic_comparison}
		\begin{tabular}{lrrrl}
			\hline
			Scenario & ECL reduction & Reduction (\%) &
			$\Delta$ objective & Paired MC 95\% interval \\
			\hline
			DAPS & 0.0004 & 0.006 & 0.000004 &
			$[-0.00029,\,0.00107]$ \\
			HWTP & 0.0508 & 0.696 & 0.003472 &
			$[0.04567,\,0.05403]$ \\
			SWUC & 0.0398 & 0.545 & 0.003046 &
			$[0.03552,\,0.04336]$ \\
			DiRe & 0.0520 & 0.711 & 0.003595 &
			$[0.04694,\,0.05547]$ \\
			\hline
		\end{tabular}
		
		\smallskip
		\parbox{0.96\textwidth}{\footnotesize
			\textit{Notes:} ECL reduction is conventional minus climate-aware analytical expected credit loss; $\Delta$ objective is climate-aware minus conventional expected surviving-interest income. Intervals describe reductions in simulated losses from 6,000 paired Gaussian Monte Carlo paths. All absolute quantities are measured on the normalized lending-budget scale.}
	\end{table}
	
	\textcolor{black}{The reductions in expected credit losses range from 0.006\% under DAPS to 0.711\% under DiRe. Under the latter scenario, the absolute reduction of approximately 0.052 budget units corresponds to 5.2 basis points of total lending exposure. Expected surviving-interest income also increases under HWTP, SWUC, and DiRe, although the changes are small relative to the portfolio-wide objective.}
	
	\textcolor{black}{The paired Monte Carlo intervals reinforce the quantitative comparisons. The intervals for expected-loss reductions exclude zero under HWTP, SWUC, and DiRe, while the DAPS interval includes zero. These intervals quantify the precision of simulated policy comparisons under the specified portfolio and model, using common latent draws to reduce the variance of policy differences.}
	
	\textcolor{black}{The results also reveal the importance of the dependence structure for portfolio risk. Under Gaussian dependence, the 99\% expected shortfall of climate-aware portfolios is approximately 16.04--16.18 budget units. Under Student-$t$ dependence with four degrees of freedom, the corresponding values increase to approximately 19.97--19.99. This substantial sensitivity of extreme losses complements the differences in expected credit losses and lending allocations. The dependence and calibration experiments are examined in greater detail in the following section.}

	\section{Discussion}
	
	\subsection{\textcolor{black}{Model scope and Future directions}}

	\textcolor{black}{Holding borrower characteristics, volatility, and terminal liability thresholds fixed isolates the consequences of alternative NGFS pathways rather than conflating them with changing calibrations. The subsequent sensitivity experiments assess how robust these comparisons are to structural and dependence assumptions.}

	\textcolor{black}{Future extensions could incorporate firm-level emissions, geographic exposures, and asset-location information to distinguish borrowers within sectors more precisely. Time-varying capital structures, climate-sensitive volatility, extreme-event shocks, and macroeconomic feedback effects would further enrich the representation of borrower adaptation and portfolio tail risk.}

	\subsection{Robustness and Sensitivity Analysis}

	\textcolor{black}{We examine how alternative dependence assumptions, structural calibrations, NGFS shock mappings, diversification constraints, and simulation choices affect estimated default probabilities, portfolio losses, and lending allocations.}

	\paragraph*{\textcolor{black}{Dependence and tail risk.}}
	\textcolor{black}{We compare Gaussian dependence with a Student-$t$ copula with four degrees of freedom, preserving obligor-level marginal default probabilities. Table \ref{tab:robustness} reports the corresponding values for climate-aware allocations. Under Gaussian dependence, ES ranges from approximately 16.04 to 16.18, compared with 19.97--19.99 under the Student-$t$ copula. Thus, heavier-tailed dependence increases estimated extreme losses by approximately 24\%, despite unchanged marginal default probabilities. This demonstrates that dependence assumptions materially affect portfolio tail-risk assessments.}

	\paragraph*{\textcolor{black}{Calibration and scenario sensitivity.}}
	\textcolor{black}{We vary production volatility and baseline productivity by $\pm20\%$, lending-rate parameters by $\pm25\%$, and climate-shock intensity by factors of $0.5$ and $2$. At fixed reference exposures and default barriers, a $\pm20\%$ change in volatility moves the DAPS mean PD from approximately 8.80\% to 18.35\%, compared with 13.95\% under the benchmark. Alternative NGFS shock mappings also change scenario rankings, illustrating the importance of the assumed climate-to-credit transmission mechanism. Additional experiments vary the bounds applied to extreme GEME3 production ratios. These results highlight the influence of structural calibration and scenario-mapping assumptions on estimated credit risk.}
	
	\paragraph*{\textcolor{black}{Diversification and numerical stability.}}
	\textcolor{black}{We re-solve the nonlinear allocation problem under individual-exposure caps $\kappa\in\{0.005,0.01,0.015,0.03,0.05\}$. Tighter caps reduce concentration and increase the number of funded obligors, while limits of 1.5\% or higher are effectively nonbinding under the benchmark calibration. The PCA approximation is also evaluated using two, five, and ten components and independent simulation seeds. Five components explain more than 99.9999\% of the model-implied systemic covariance variance. Analytical and Monte Carlo default probabilities agree within four simulation standard errors for all obligors examined.}
	
	\begin{table}[htbp]
		\centering
		\footnotesize
		\setlength{\tabcolsep}{4pt}
		\caption{Dependence and diversification sensitivity}
		\label{tab:robustness}
		\begin{tabular}{lrrrr}
			\hline
			Scenario & Gaussian ES$_{99}$ & $t$-copula ES$_{99}$
			& $N_+$ (0.5\% cap) & $N_+$ (5\% cap) \\
			\hline
			DAPS & 16.184 & 19.990 & 274 & 237 \\
			HWTP & 16.037 & 19.973 & 278 & 238 \\
			SWUC & 16.062 & 19.977 & 278 & 238 \\
			DiRe & 16.044 & 19.975 & 278 & 238 \\
			\hline
		\end{tabular}
	\end{table}

	\textcolor{black}{Taken together, these experiments strengthen the numerical and methodological foundations of the framework and clarify the influence of alternative modelling assumptions on credit-risk and lending outcomes. Agreement between analytical and simulated default probabilities supports the consistency of the implementation, while sensitivity to dependence structures, structural calibration, and diversification constraints identifies economically important sources of model uncertainty. The findings demonstrate the value of evaluating climate-related credit risk across alternative specifications and provide a foundation for further applications using institution-specific banking data.}

	\section{Practical Implementation and Supervisory Use}

	\textcolor{black}{The framework provides a basis for incorporating NGFS climate scenarios into existing bank credit-risk and portfolio-management processes. An institutional implementation would combine loan-level exposures, borrower financial information, sector classifications, lending spreads, collateral characteristics, and maturities with internally calibrated default probabilities and loss-given-default estimates. Mapping NGFS projections to these exposures would generate climate-conditioned measures of obligor creditworthiness, portfolio losses, and lending allocation outcomes. Consistent sector classifications and institution-specific structural calibration would support the integration of scenario information with existing credit-risk systems.}
	
	\textcolor{black}{The numerical implementation evaluates 1,000 obligors using 6,000 Monte Carlo draws and constrained nonlinear lending optimization. Analytical marginal default probabilities and covariance-based principal component reduction improve computational efficiency, while simulation captures dependence and portfolio loss distributions. Applications to larger lending portfolios could employ parallel computation, efficient numerical routines, and systematic convergence diagnostics to assess scalability and the stability of optimization results.}
	
	\textcolor{black}{Model governance would involve documented scenario selection, parameter calibration, data lineage, independent validation, and sensitivity analysis. Validation against observed credit outcomes would support institution-specific assessment of predictive performance, while alternative dependence structures and scenario-transmission assumptions would inform model-risk evaluation. Since NGFS pathways represent conditional economic developments, their use in regulatory and accounting applications requires appropriate treatment of scenario probabilities, forecast horizons, and relevant credit-risk parameters. Optimization diagnostics would also assess convergence and sensitivity to initial allocations.}
	
	\textcolor{black}{For bank risk managers, the framework supports comparisons of sectoral credit exposures, expected losses, lending income, and portfolio concentration across alternative climate pathways. These measures can inform credit portfolio reviews, climate stress testing, and lending policy decisions. For supervisors, the analysis provides a transparent basis for examining scenario sensitivity, concentration risk, and the effects of alternative dependence assumptions on extreme portfolio losses. Applications within ICAAP, IFRS~9, and internal-ratings-based (IRB) capital processes would incorporate institution-specific PD, LGD, and exposure-at-default calibration, together with the relevant accounting or regulatory requirements and governance arrangements \cite{ifrs9,bcbs2022,eba2025management,eba2025scenario}.}

	\section{Conclusion}

\textcolor{black}{This study links NGFS short-term scenario shocks to structural terminal default probabilities and lending decisions. Sectoral shocks affect borrower production and creditworthiness, while the lending exposure itself alters default risk. This interaction makes allocation endogenous and provides a basis for comparing lending income, credit losses, and portfolio risk across climate pathways.}

	\textcolor{black}{The numerical results reveal substantial heterogeneity in the distribution of climate-related credit risk. Although mean five-year default probabilities range from approximately 13.88\% to 13.95\% across scenarios, differences in sectoral creditworthiness generate meaningful changes in lending allocations. Relative to conventional lending, climate-aware allocation reduces expected credit losses by approximately 0.55--0.71\% under HWTP, SWUC, and DiRe, accompanied by small improvements in expected surviving-interest income. Comparisons with equal-weight and fixed-PD allocations highlight the interaction between lending concentration, expected returns, and endogenous default risk. Furthermore, the Student-$t$ dependence specification produces approximately 24\% higher portfolio expected shortfall than Gaussian dependence, underscoring the importance of joint-default assumptions in assessing extreme losses.}

	\textcolor{black}{These findings demonstrate the relevance of combining scenario-dependent credit-risk assessment with lending optimization and portfolio loss analysis. For financial institutions, the framework provides a systematic basis for identifying sectoral vulnerabilities, evaluating lending policies, and examining portfolio resilience under alternative climate pathways. Its implementation with institution-specific exposure, borrower, and credit-performance data would support empirical calibration and validation, while extensions incorporating dynamic lending decisions, borrower adaptation, and macroeconomic feedback effects would further strengthen its application to climate-related financial risk management.}

	%% This adds a line for the Bibliography in the Table of Contents.
	\addcontentsline{toc}{chapter}{Bibliography}
	%% ***   Set the bibliography style.   ***
	\bibliographystyle{plain} % (change according to your preference)
	%%% ***   Set the bibliography file.   ***
	%\bibliography{DSASMScProject}{}
	%% ***   NOTE   ***
	%% If you don't use bibliography files, comment out the previous line
	%% and use \begin{thebibliography}...\end{thebibliography}.  (In that
	%% case, you should probably put the bibliography in a separate file
	%% and \include or \input it here).

\end{document}